\documentclass{article}

\usepackage{iclr2025_conference,times}
\usepackage{graphicx}
\usepackage[hidelinks]{hyperref}
\usepackage{url}

\title{A Quality of Mercy is Not Trained:\\ The Imagined vs.\ the Practiced in Healthcare Process-Specialized AI Development\thanks{Accepted to SMASH 2025.}}

\author{Anand Bhardwaj \\
Desautels Faculty of Management\\
McGill University\\
\texttt{anand.bhardwaj@mail.mcgill.ca}
\And
Samer Faraj \\
Desautels Faculty of Management\\
McGill University\\
\texttt{samer.faraj@mcgill.ca}
}

\iclrfinalcopy
\begin{document}
\maketitle

\begin{abstract}
In high-stakes organizational contexts like healthcare, artificial intelligence (AI) systems are increasingly being designed to augment complex coordination tasks. This paper investigates how the ethical stakes of such systems are shaped not only by their outputs but by their epistemic framings: what aspects of work they represent, and what they exclude. Drawing on an embedded study of AI development for operating room (OR) scheduling at a Canadian hospital, we compare ``scheduling as imagined'' in the AI design process: rule-bound, predictable, and surgeon-centric, with ``scheduling as practiced'' as a fluid, patient-facing coordination process involving ethical discretion. We show how early representational decisions narrowed what the AI could support, resulting in \textit{epistemic foreclosure}: the premature exclusion of key ethical dimensions from system design. Our findings surface the moral consequences of abstraction and call for a more situated approach to designing healthcare process-specialized artificial intelligence systems.
\end{abstract}

\section{Introduction}

Artificial intelligence (AI) is increasingly being embedded into high-stakes organizational settings under the promise of optimizing complex coordination processes. In healthcare, AI systems are now being deployed to help schedule surgeries, triage patients, and manage beds—tasks traditionally reliant on human judgment, discretion, and lived relationships with patients. As this wave of digital transformation proceeds, researchers and practitioners alike have raised questions about what such systems recognize, represent, and exclude \citep{bailey2022relational, kellogg2020algorithms, pachidi2021algorithms}.

Designing novel AI systems, especially those intended to augment complex coordination processes, rarely begins from a stable foundation. Early-stage development often unfolds under conditions of epistemic and institutional uncertainty: it may be unclear what specific problem the system is solving, what broader strategic goal it serves, which organizational process it will participate in, or how it will integrate with existing workflows. In such contexts, design decisions are doubly consequential: they shape the contours of the system in the present, while setting into motion path dependencies that structure what the technology can become. These early choices are rarely neutral. They encode particular imaginations of process, elevate certain forms of knowledge, and preclude others, often before such trade-offs are visible or deliberate. These moments, where assumptions must be made despite uncertainty, give rise to what we call moments of knowing: provisional resolutions about what a process is, what matters in it, and how it ought to be supported. As scholars of technological development have long observed, early representational moves often “close” interpretive flexibility, stabilizing the meaning and use of the system prematurely \citep{pinch1984social}.

Building on this insight, prior research in organizational theory and information systems has shown how AI systems reconfigure decision-making by formalizing tacit practices and privileging codified forms of knowledge \citep{faraj2018working, raisch2020automation}. In healthcare, such codification is particularly fraught: what appears “efficient” to an algorithm may conflict with evolving patient conditions, clinician judgment, or the ethical imperative to do no harm \citep{borry2020rulebreaking, faraj2006coordination}. Operating room (OR) scheduling exemplifies this tension. While often described as a logistical task, it is in fact a complex coordination process—one that spans multiple occupational domains, accommodates situational ethics, and relies heavily on local, situated knowledge.

This study draws on an embedded field study at the Goodman University Hospital, a Canadian tertiary-care hospital co-developing a predictive AI system for OR scheduling with a local developer. The system's design intent is to reduce "wasted" OR time by identifying optimal OR schedules based on accurately predicting the duration of surgical procedures, identifying risk of procedures extending beyond scheduled duration, and enforcing multiple cascading prioritization logics. Yet early in the design process, we observed divergences in what the system was "seeing" as the essential knowledge basis of scheduling, from the involved ethical discretion and patient-facing reasoning practiced by human schedulers. As one associate director noted: “Every day, I’m deciding: \emph{you} will have your surgery canceled, and \emph{you} will have it today.”

We argue that AI systems do not simply support scheduling: they instantiate a particular imagination of what scheduling is and how it "should" be done. In the case we examine, this meant foregrounding structured administrative and clinical codes while sidelining clerical judgment, interpersonal discretion, and tacit forms of ethical reasoning. We conceptualize this through the construct of \textit{epistemic foreclosure}: a narrowing of epistemic possibilities produced by early representational choices in AI system design. This framing draws from and contributes to relational views of knowledge and sociotechnical systems, emphasizing how tools emerge not in isolation but through contested developmental processes entangled with local knowings and institutional logics \citep{bailey2022relational, faraj2022uncertainty, waardenburg2022brokerage}.

By comparing “scheduling as imagined” in the AI design process with “scheduling as practiced” by clerks, nurses, and managers, we surface how epistemic exclusions shape not only what the system does, but what it fails to notice, support, or care about.

\section{Methods}

This study draws on an embedded, qualitative field study at the Goodman, a Canadian tertiary-care hospital, where a predictive scheduling tool was being co-developed to optimize the allocation of surgical cases to operating rooms (ORs). Rather than treating AI implementation as a bounded event, we approach the field site as an entanglement of two temporally and materially distinct processes: the ongoing, institutionalized process of OR scheduling, and the one-time, project-based process of AI development. 

This framing draws inspiration from Karen Barad’s concept of \textit{entanglement} as not merely connection or interaction between separate entities, but the constitutive intra-action through which actors, artifacts, and ways of knowing emerge and become intelligible in relation to each other \citep{barad2007meeting}. At Goodman, this entanglement gave rise to specific moments where AI developers, clerks, nurses, and managers were called upon to articulate, justify, or revise their assumptions about how scheduling works, and how it should be represented. These moments were both epistemic and ethical, as they revealed tensions between formal rules and discretionary care.

Data collection spanned multiple perspectives across these two processes. We attended 23 design and coordination meetings (27.5 hours) involving clinicians, hospital managers, and developers; shadowed OR programming clerks over 22 sessions (82 hours); and conducted 36 formal and 20 informal interviews, totaling 35.4 hours. We also reviewed 20 internal design documents—process maps, prioritization rules, data schemas, and system requirement memos—totaling 478 pages.

Our analytic focus was on how these entangled processes produced what we term “moments of knowing”: situated episodes where the assumptions built into the AI (scheduling as imagined) came into contact with the epistemic demands of everyday scheduling (scheduling as practiced). We identified these moments inductively, often beginning from participant-flagged episodes of breakdown, mismatch, or discomfort—what Suchman might describe as “trouble” in sociotechnical systems \citep{suchman2007human}. These moments were then traced across documents, meetings, and fieldnotes to understand how they were interpreted, resolved, or foreclosed.

Our analytic approach is processual and relational: We do not treat technologies, rules, or roles as stable entities, but as provisional accomplishments shaped through their intra-actions with local practices and institutional priorities \citep{bailey2022relational, faraj2018working, faraj2022uncertainty}. In attending to how the AI system and the scheduling process modified each other’s contours over time, we seek to surface the epistemic and ethical consequences of design choices often treated as technical or administrative.

\section{Findings: The ethics of exclusion}

The AI development effort at Goodman was anchored in a particular imagination of what operating room (OR) scheduling entailed. This imagination, primarily articulated by the surgeons and project sponsors who participated in development, framed scheduling as a rule-based administrative process that could be optimized using existing data. Yet ethnographic observation and interviews revealed that everyday scheduling was a discretionary and situated process, routinely improvised by clerks, nurses (not included in the development process) in response to emergent conditions.

\subsection{Scheduling as Imagined}

In design meetings, surgeons described scheduling as a matter of enforcing pre-existing triage rules and ensuring that clerks followed surgeon directives. Several surgeons expressed skepticism about the epistemic contributions of clerks, viewing them primarily as implementers. One surgeon noted, “They just fill the slots. We’ve already made the decisions.”

The AI system was designed accordingly. It encoded surgeon-determined prioritization categories, surgeon-identified drivers of surgical duration in case data, and administrator-directed waitlist order as primary determinants. Patient-facing considerations—such as anxiety, readiness, or socioeconomic disruptions—were treated as noise. The goal was to remove discretion from scheduling and replace it with consistency, transparency, and efficiency.

This vision rendered invisible the forms of situated knowledge that clerks and nurses mobilized daily. It also depersonalized the moral labor of triage. 

\subsection{Scheduling as Practiced}

Contrary to the system’s assumptions, OR scheduling at Goodman unfolded as a fluid, interpretive practice. This practice was not merely reactive, but entailed anticipatory coordination, grounded in familiarity with organizational dynamics, interpersonal relationships, and unwritten priorities. Drawing from extended shadowing and interviews, we identified four recurring moments that structured this work:

\textbf{1. Interpreting the Requisition.}  
Clerks did not simply enter surgical requests into a queue. This is a step that involves both interpreting surgeon shorthand and other context cues, and assembling additional information required for scheduling, beyond what is in the form. Clerks often paused to assess whether a patient was truly ready for surgery, based on informal cues in documentation, recent cancellations, or notes from the pre-op team. This judgment was relational and practical, not formalized in any system field.

\textbf{2. Anticipating Surgeon Behavior.}  
Clerks routinely accounted for surgeon-specific patterns: who overbooks, who finishes early, who insists on particular sequences. This anticipatory work was crucial to building viable schedules and avoiding cascading disruptions.

\textbf{3. Managing Patient Cancellations.}  
Cancellations were not binary. Clerks distinguished between cancellations with minimal patient impact and those with high emotional or logistical cost (e.g., long-distance travel, childcare arrangements). These distinctions informed how clerks rescheduled or prioritized subsequent cases.

\textbf{4. Negotiating with the OR team.}  
The schedule was constantly in flux. Clerks adjusted for emergency add-ons, equipment failures, and anesthesiologist availability. These changes required improvisation, not just rule-following—and they often called upon clerks’ embedded knowledge of patient histories and staff dynamics.

Across these moments, scheduling appeared not as a problem to be solved, but as an ongoing negotiation of fairness, feasibility, and care. 

\subsection{From Situated Discretion to an Optimization Problem}

Despite these complexities, the AI system recast scheduling as a tractable optimization problem. This translation produced friction at key junctures where the epistemic assumptions of the system came into contact with everyday work. Here we identify examples that surfaced ethical tensions, representational exclusions, and subtle forms of epistemic foreclosure.

\textbf{Tension 1: The Letter vs. Spirit of the Law.}  
The Goodman also is a tier 1 trauma center, which means scheduled surgeries can be cancelled to accommodate unscheduled, emergency procedures. To address this, the AI system encoded a rule to prioritize “Previously Cancelled” (PC) patients. However, the system only captured cancellations that occurred on the day of surgery. This excluded patients whose cancellations were equally disruptive but logged the week or night before the day of surgery. As one clerk remarked, “He chartered a plane to come in and got canceled. He didn’t get PC.” The system’s rule failed to capture the ethical intent behind the PC label.

\textbf{Tension 2: Conflicting Logics of Priority.}  
In development meetings, it emerged that multiple logics of prioritization exist, often in conflict with each other. These conflicts, invisible thanks to the clerks' efforts to negotiate and compromise, was made visible when various stakeholders articulated their own particular version of the logic. According to administrators, the dominant logic of prioritization is wait-time, a visible problem in the regional healthcare system. Yet this conflicted with clinical urgency. A nurse manager asked: “Do you want to put this one first, because [the patient] has been waiting almost a year, or do you want to put the cancer first?” 

\textbf{Tension 3: Coding to the System.}  
The principal source of "labeling" of data meant for supervised machine learning to predict surgical duration was the surgeon-entered surgical code attached to each past surgery (and its actual duration).  However, this labeling was done prior to the system development project, and for different pragmatic reasons than achieving an accurate predictive system. Two key instances stand out. First, surgeons adapted their behavior to game the system—sometimes using surgical codes that equip the OR as they wish, but with a shorter average  duration. A clerk noted, “She always puts 2.5 hours, but it’s always at least 3.5 or 4.” Second, different political or pragmatic practices across surgical specialties (e.g. some exclude prep time in the recorded duration of a case, some exclude anaesthesia time), added noise to the training data.

Together, these moments demonstrate how the transformation of situated discretion into a computational logic narrowed the ethical and epistemic space of scheduling. Rather than supporting coordination, the system redefined what counted as coordination, transforming a moral and relational practice into a computational abstraction.

\subsection{From Coordination to Codification}

The exclusions observed in system design arose from an underlying epistemic misalignment. Scheduling-as-practice lacks a stable ground truth: it is a situated, interpretive process that unfolds through discretion, anticipation, and ethical judgment. Different actors like clerks, nurses, surgeons, and managers draw on distinct, incommensurate forms of knowing. This multiplicity is not a breakdown to be resolved, but a feature of the coordination process itself. It is this interplay of perspectives that allows the scheduling system to remain flexible, responsive, and just.

Yet the AI system was built on the premise that this process could be fully represented, codified, and optimized. This shift—from coordination-as-practiced to coordination-as-imagined imposed a fixed logic onto a fluid practice. Early design decisions constrained what the AI could meaningfully see, reason about, or support. We conceptualize this constraint as \textit{epistemic foreclosure}: the premature narrowing of moral and practical possibilities through representational choices that rendered some forms of knowing illegible. Decisions that were once shared, discussed, or deliberated became black-boxed outputs, replacing moral discretion with procedural certainty, and tacit negotiation with algorithmic finality.

\section{Discussion and Contributions}

This study contributes to research on AI ethics and sociotechnical systems by highlighting a foundational challenge in applying AI to complex organizational settings: there may not be a dominant "ground truth" towards which such systems can converge. In domains like healthcare, processes such as scheduling are not fixed rule sets but negotiated practices—enacted through situated judgment, ethical discretion, and evolving constraints. When AI systems are built to optimize such processes, they must navigate this absence of stable reference points by participating in decisions about what the process is, what matters within it, and how it should be supported.

This epistemic indeterminacy makes early design choices especially consequential. What gets represented, and what gets excluded, shapes not only what the system can do, but what ways of knowing are rendered intelligible or irrelevant. We argue for a situated ethics of AI—one that attends not only to algorithmic fairness or output accuracy, but to how knowledge is enacted, negotiated, and stabilized. Especially in healthcare, designers must ask not only whether a system works, but \textit{whose work} it supports, and at what cost. Designing AI for such contexts demands more than encoding knowledge; it requires building systems that can remain open to dialogic revision, responsive to shifting knowings across roles, values, and time.

\bibliographystyle{iclr2025_conference}
\bibliography{mercy2025}

\end{document}